\documentclass[showpacs,aps,twocolumn,superscriptaddress]{revtex4}%
\usepackage{graphicx}
\usepackage{pdfpages}
\usepackage{dcolumn}
\usepackage{bm}
\usepackage{amsmath}
\usepackage{fancybox}
\usepackage{amsfonts}
\usepackage{amssymb}

\newcommand{\s}{\sum\limits}

\newcommand{\pa}{\partial}

\newcommand{\be}{\begin{equation}}
\newcommand{\e}{\end{equation}}
\newcommand{\beml}{\begin{subequations}}
\newcommand{\eml}{\end{subequations}}
\newcommand{\beq}{\begin{eqnarray}}
\newcommand{\eq}{\end{eqnarray}}
\newcommand{\ba}{\begin{array}}
\newcommand{\ea}{\end{array}}
\newcommand{\bpm}{\begin{pmatrix}}
\newcommand{\epm}{\end{pmatrix}}
\newcommand{\bc}{\begin{cases}}
\newcommand{\ec}{\end{cases}}
\newcommand{\lt}{\left}
\newcommand{\rt}{\right}
\newcommand{\n}{\nonumber}
\newcommand{\la}{\langle}
\newcommand{\ra}{\rangle}
\newcommand{\ep}{\varepsilon}

\newcommand{\bb}{\boldsymbol}
\newcommand{\h}{^\dagger}
\newcommand{\0}{^\phantom{\dagger}}

\newcommand{\nimp}{n_{\textrm{imp}}}

\DeclareMathOperator{\tr}{Tr}

\DeclareMathOperator{\im}{Im}
\DeclareMathOperator{\re}{Re}

\begin{document}
\title{Theory of light-induced effective magnetic field in Rashba ferromagnets}
\author{Alireza Qaiumzadeh}
\affiliation{Institute for Molecules and Materials, Radboud University, 6525 AJ Nijmegen, The Netherlands}
%\affiliation{Department of Physics, Institute for Advanced Studies in Basic Sciences (IASBS), Zanjan 45137-66731, Iran}
\author{Mikhail Titov}
\affiliation{Institute for Molecules and Materials, Radboud University, 6525 AJ Nijmegen, The Netherlands}
\begin{abstract}
Motivated by recent experiments on all-optical magnetization reversal in conductive ferromagnetic thin films we use non-equilibrium formalism to calculate the effective magnetic field induced in a Rashba ferromagnet by a short laser pulse. The main contribution to the effect originates in the direct optical transitions between spin-split sub-bands. The resulting effective magnetic field is inversely proportional to the impurity scattering rate and can reach the amplitude of a few Tesla in the systems like Co/Pt bi-layers. We show that the total light-induced effective magnetic field in ferromagnetic systems is the sum of two contributions: a helicity dependent term, which is an even function of magnetization, and a helicity independent term, which is an odd function of magnetization. The primary role of the spin-orbit interaction is to widen the frequency range for direct optical transitions.
\end{abstract}
\pacs{78.20.Bh, 78.20.Ls, 75.70.Tj, 42.65.-k}
\date{\today}
\maketitle

\section{Introduction}

The discovery of all-optical magnetization reversal \cite{Kimel2005,Stanciu2007} has been broadly recognized as an important step forward to ultra-fast opto-magnetic data recording and processing. The phenomenon has opened up a topic in the field of spintronics that is challenging from both experimental and theoretical points of view \cite{IFE-review,IFE-review0,IFE-review1}.

Ultrafast spin dynamics in ferromagnetic nickel has been demonstrated by Beaurepaire \textit{et al.} in 1996.\cite{Beaurepaire} The experiment triggered a lot of interest to the possibility of controlling magnetization by optical means. A decade later, Rasing and co-workers \cite{Kimel2005,Stanciu2007,IFE-review} have indeed demonstrated that a sub-picosecond laser pulse of circularly polarized light may reverse magnetization direction in metallic ferrimagnets such as GdFeCo \cite{GdFeCo1,GdFeCo10,GdFeCo2}, TbCo \cite{TbCo} as well as in synthetic ferrimagnet compounds \cite{synthetic-fm}. Despite the experimental significance of the effect its microscopic origin is still debated \cite{Pershan,Shen,IFE-plasma1,IFE-plasma2,Blugel1,Blugel2,Alireza1,Oppeneer,haet-switch,haet-switch1}.

Very recently all-optical switching in ferromagnetic Co/Pt bilayer irradiated by a $100$\,fs laser pulse has been reported \cite{exp-metalic-IFE}. Motivated by this experiment we analyze the effective magnetic field induced by polarized light in a two-dimensional (2D) ferromagnet with Rashba spin-orbit coupling. We find that the strong spin-orbit interaction in the system \cite{co-pt,Alireza2} opens up a wide frequency window for direct optical transitions that induce the effective magnetic fields up to few Tesla. We, therefore, argue that the effect described can be central for understanding magnetization switching in Co/Pt bilayers and similar heterostructures.

Below we consider a 2D ferromagnet subject to a polarized light propagating along the magnetization direction that is normal to the $xy$-plane of the ferromagnet. For the case of a circularly polarized light there is no anisotropy in the $xy$-plane, hence the effective magnetic field may only be induced in the $z$ direction. We distinguish odd and even components of the effective magnetic field with respect to the helicity of light that are regarded as inverse Faraday (IF) and inverse Cotton-Mouton (ICM) contributions, respectively \cite{IFE-review0,Shen}. A linearly polarized light leads in addition to a small magnetic field component in the direction of light polarization.

We calculate the largest, so-called kinematic, contribution to the effective magnetic field that is facilitated by the direct optical transitions between spin-split sub-bands. The stronger the spin-orbit interaction the wider the frequency window for such direct optical transitions. Modeling the dynamics of magnetization switching remains, however, beyond the scope of the paper.

The rest of the paper is organized as follows. In Sec.~\ref{sec:model} we introduce the theoretical model and the technique. In Sec.~\ref{sec:myfield} we calculate the effective magnetic field induced by polarized light up to the second order with respect to the light intensity. In Sec.~\ref{sec:conductivity} the optical conductivity and its relation to the absorption in this system is discussed. We conclude in Sec.~\ref{sec:conclude}.

\section{Model Hamiltonian and formalism}\label{sec:model}
\subsection{System Hamiltonian}

Rashba spin-orbit interaction originates in inversion symmetry breaking due to the electron 2D confinement. In heavy-metal/ferromagnet bilayers such as Co/Pt the spin-orbit coupling is particularly strong \cite{co-pt,Rashba-SOI,review-Rashba1,review-Rashba2,k-linear}. Conduction electrons in such systems can be qualitatively described by the Bychkov-Rashba Hamiltonian \cite{Bychkov84},
\be
\label{model}
\hat{H}=\hat{H}_0+\hat{V}(\bb{r}),\quad \hat{H}_0=\frac{\hat{\bb{p}}^2}{2m}+\alpha_\textrm{R} (\hat{\bb{p}}\times\bb{\sigma})_z +M\sigma_z,\;\;
\e
where $\hat{V}(\bb{r})$ is a disorder potential, $\hat{\bb{p}}$ is the momentum operator in two dimensions, $m$ is the effective electron mass, $\alpha_\textrm{R}$ quantifies the strength of spin-orbit interaction, $\bb{\sigma}$ is the vector of Pauli matrices representing spin operators of itinerant electrons, and $M$ is the ferromagnetic exchange energy for magnetization along the $z$ direction that is normal to the 2D plane.

%%%%%%%%%%%%%%%%%%%%%%%%%%%%
%%%% fig:spectrum
%%%%%%%%%%%%%%%%%%%%%%%%%%%%
\begin{figure}[bt]
\includegraphics[width=0.75\columnwidth]{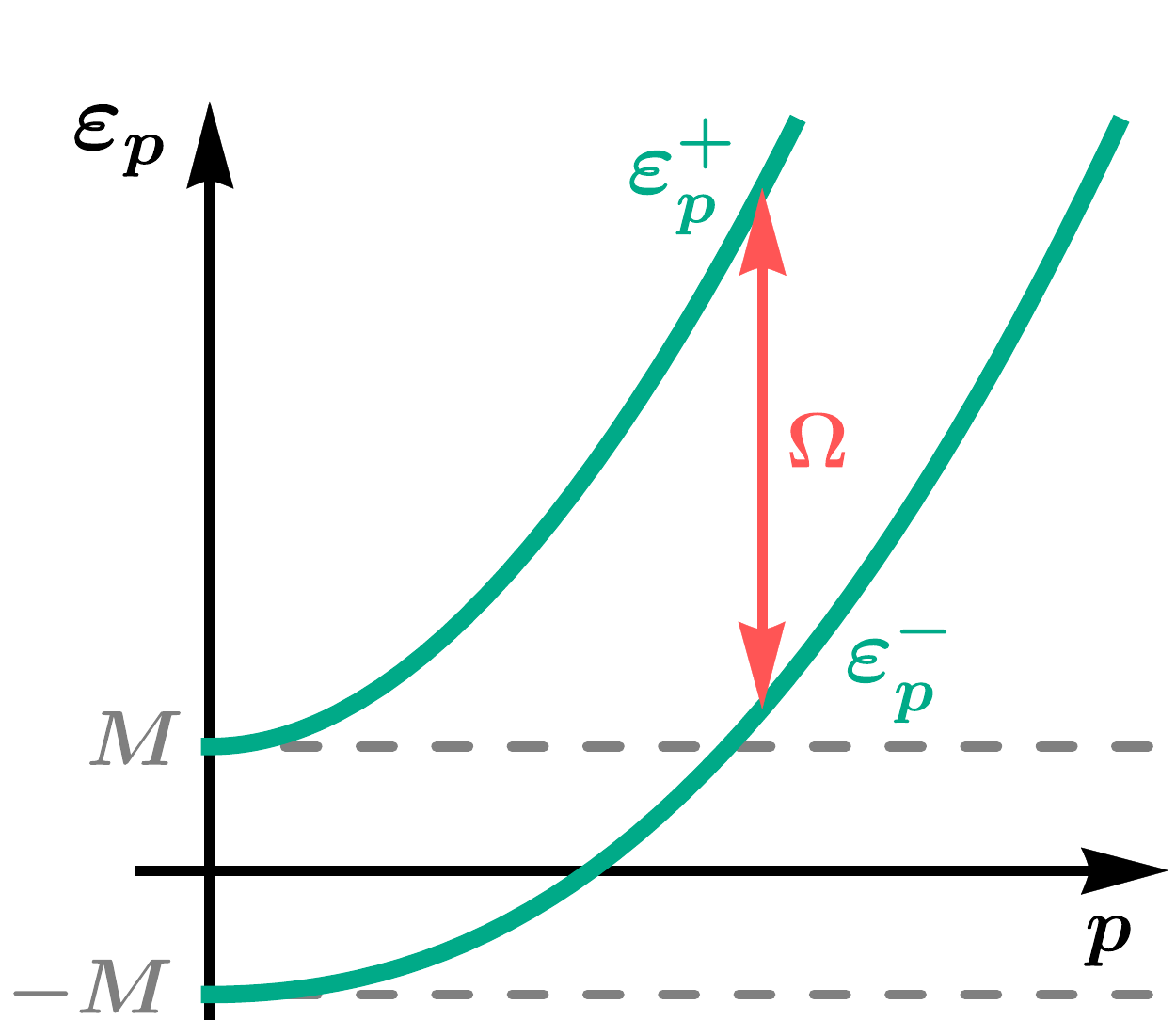}
\caption{The spectrum of model (\ref{model}) and the direct optical transition corresponding to the frequency $\Omega$ that fulfils the condition of Eq.~(\ref{condition}).}
\label{fig:spectrum}
\end{figure}
%%%%%%%%%%%%%%%%%%%%%%%%%%%%%

The effect of light is introduced by the Peierls substitution $\hat{\bb{p}}\to \hat{\bb{p}}-e \bb{A}(t)$, where $\bb{A}(t)$ is the time-dependent vector potential and $e$ is the electron charge. The corresponding electric field is given by $\bb{\mathcal{E}}(t)=-\pa \bb{A}(t)/\pa t$. Then the total Hamiltonian is replaced by $\hat{H}\to \hat{H}+\hat{\bb{j}}(\hat{\bb{p}}) \!\cdot\! \bb{A}(t) + e^2 A(t)^2/2m$, where $\hat{\bb{j}}(\hat{\bb{p}})=e\lt(\hat{\bb{p}}/m+ \alpha_\textrm{R}\hat{\bb{z}}\times\bb{\sigma}\rt)$ is the current operator. Throughout the paper we use the units with $c=\hbar=1$.

In the case of light propagating in $z$ direction, i.\,e.\, perpendicular to the $2D$ plane of electron gas, the electric field component of laser pulse can be described by a spatially uniform electric field $\bb{\mathcal{E}}(t)=\mathcal{E}_0\bb{e}_\lambda e^{-i\Omega t}$ where $\mathcal{E}_0$ is the amplitude of the electric field and $\Omega$ is the frequency of light. The unit vector $\bb{e}_\lambda=(\hat{\bb{x}}+i\lambda\hat{\bb{y}})/\sqrt{1+\lambda^2}$ defines the light polarization vector: $\lambda=\pm 1$ for left (right) circularly polarized light and  $\lambda=0$ for light polarized in the $x$ direction. In what follows we regard the parameter $\lambda$ as the light helicity. The corresponding vector potential is given by $\bb{A}(t)=\mathcal{E}_0\bb{e}_\lambda e^{-i\Omega t}/i\Omega$.

The spectrum of the free Hamiltonian $\hat{H}_0$ consists of two spin-split sub-bands $\ep^\pm_{\bb{p}}= p^2/2m\pm\sqrt{\alpha_\textrm{R}^2p^2+M^2}$ that are depicted schematically in Fig.~\ref{fig:spectrum}. The direct optical transitions between the sub-bands correspond to $\Omega=\ep^+_{\bb{p}}-\ep^-_{\bb{p}}$ provided that the state with $\ep=\ep_+$ is empty while the state with $\ep=\ep_-$ is filled, i.\,e.\,$\ep^-_{\bb{p}}<\ep_\textrm{F}<\ep^+_{\bb{p}}$, where $\ep_\textrm{F}$ stands for the Fermi energy. This purely kinematic condition can be rewritten as
\be
\label{condition}
\omega_0-\Delta < \Omega/2 < \omega_0+\Delta,
\e
where $\Delta=m\alpha_\textrm{R}^2$ and $\omega_0=\sqrt{\Delta^2+2\Delta\ep_\textrm{F}+M^2}$.  Throughout this paper we focus on the situation when all electronic states involved in optical transition correspond to the energies in the upper part of the spectrum, $ \varepsilon_p > M $, as shown in Fig.~\ref{fig:spectrum}. This is the typical situation in ferromagnetic metals. In the model considered, the condition corresponds to the additional constraint $\ep_\textrm{F}>M+\Omega$ for the applicability of our results, which is equivalent to $\Omega>2(M+2\Delta)$.

In the absence of spin-orbit interaction, $\Delta=0$, the condition (\ref{condition}) is limited to $\Omega=2 M$ which is the consequence of the fact that the sub-band splitting is equal to $2M$ for all momenta. For strong spin-orbit interaction the condition (\ref{condition}) is, however, not that restrictive. For example, for the following material parameters $\ep_\textrm{F}=3.2$\,eV, $M=0.4$\,eV, and $\Delta=100$\,meV, the condition (\ref{condition}) is fulfilled for $1.6$\,eV$<\Omega< 2$\,eV. This is indeed a sufficiently large interval within optical frequency range.

In the model (\ref{model}), we consider a spin dependent and short range disorder potential as \cite{spin-dependent-dis}
\be
\hat{V}({\bb{r}})=V \sum_i \delta({\bb{r}}-{\bb{R}}_i), \quad V=\bpm V_+& 0 \\0 & V_- \epm,
\e
where the points ${\bb{R}}_i$ specify random impurity positions and $V_{\pm}$ is the characteristic impurity strength for the spin along (or opposite to) the magnetization direction.

\begin{figure}[bt]
\includegraphics[width=0.9\columnwidth]{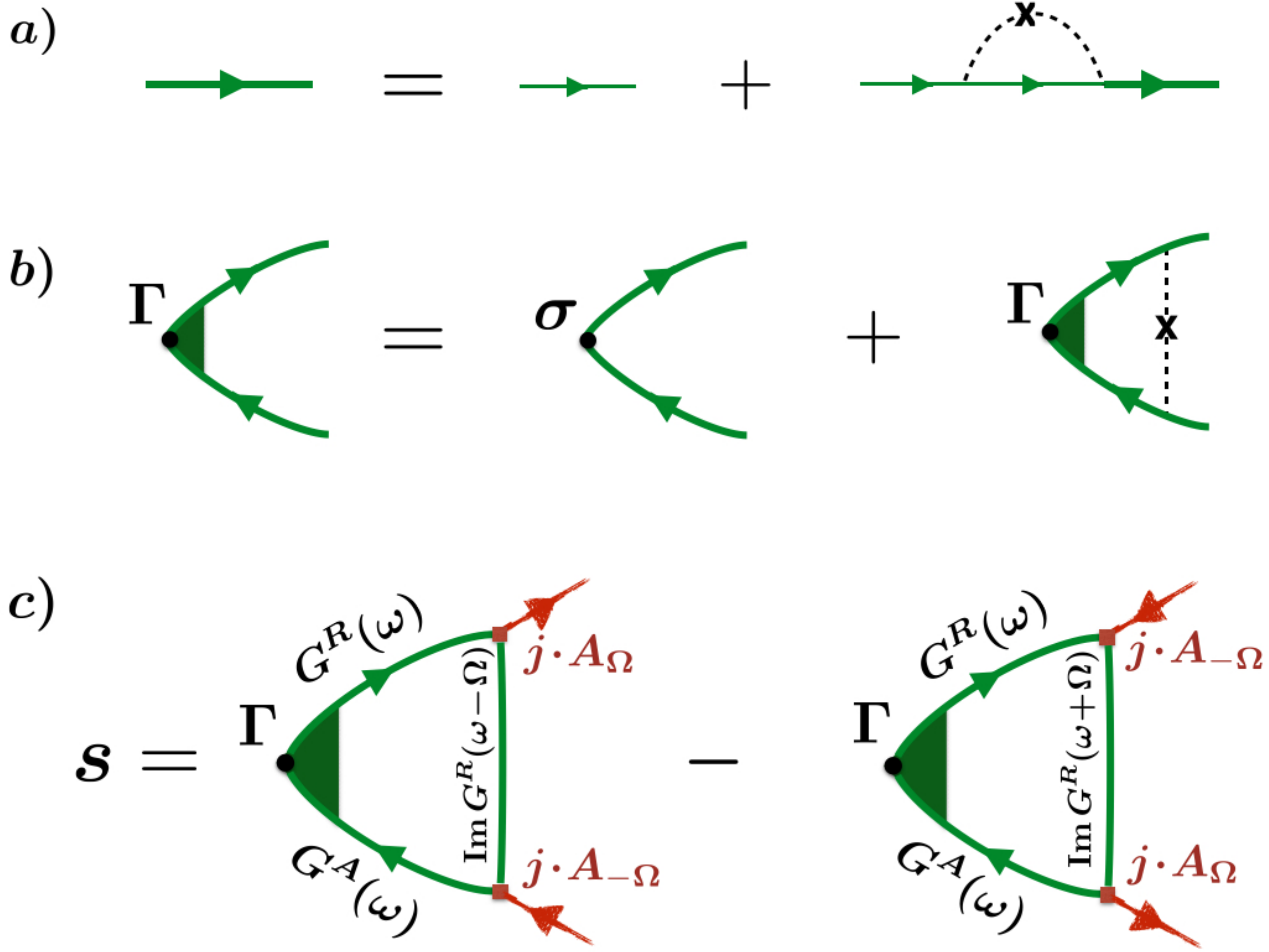}
\caption{(a) Feynman diagrams illustrating Born approximation that is used to derive the expression for the averaged Green's function in Eq.~(\ref{GR}). (b) Diagrammatic representation of Eq.~(\ref{eq}) on the vertex $\Gamma$ that describes the electron diffusion. (c) Representation of the dc non-equilibrium spin polarization due to direct optical transitions as given by Eq.~(\ref{effective-field}).}
\label{fig:field}
\end{figure}

For electrons with energies in the upper sub-band, $\ep_p > M$, we define the corresponding scattering rates $\gamma_{\pm}=\pi \nimp \nu V^2_{\pm}$, $\gamma=(\gamma_++\gamma_-)/2=1/2\tau$, where $\nimp$ is the 2D impurity concentration, and $\nu=m / 2 \pi$ is the density of states per spin. We consider the limit of weak gaussian disorder, that formally corresponds to $V_\pm\to 0$ and $\nimp\to \infty$ such that $\gamma\ll (\ep_F-M)$. In this limit the averaging over disorder realizations is expressed in terms of the single corelator
\be
\lt\la V(\bb{r})V(\bb{r}') \rt\ra-\lt\la V(\bb{r}) \rt\ra^2 =  \nimp V^2\, \delta(\bb{r}-\bb{r}'),
\e
that is depicted by the impurity line (dashed) in the Fig.~\ref{fig:field}b.

\subsection{Expression for the effective magnetic field}

The effective magnetic field is calculated to second order with respect to the electric field component of light $\bb{\mathcal{E}}$.  The contribution of the second order perturbing potential in the total Hamiltonian, $e^2 A(t)^2/2m$, to the effective magnetic field is negligible as explained in the next section. Thus, it is sufficient to construct the second order perturbation theory with respect to the linear perturbing potential $\hat{U}(\hat{\bb{p}},t)=\hat{\bb{j}}(\hat{{\bb{p}}})\!\cdot\! \bb{A}(t)$. The calculation is performed with the help of non-equilibrium perturbation theory using Keldysh formalism.

In the non-equilibrium quantum theory the effective magnetic $\bb{\mathcal{H}}$-field induced by light can be written as \cite{Edelstein,Tatara1,Murakami}
\be
\label{verygeneral}
\bb{\mathcal{H}} = -\frac{g \mu_B}{d} \bb{s},\qquad  s_a=-\frac{i}{2} \tr \lt\la \sigma_a \mathcal{G}^{<}({\bb{r}},t;{\bb{r}},t) \rt\ra,
\e
where $\mu_B$ is the Bohr magneton, $g$ is the electron effective Land\'{e} $g$-factor, and $d$ is the effective sample thickness (given by the light penetration depth for thick samples). The brackets $\la\dots \ra$ stand for disorder averaging, $a=\{x,y,z\}$, and $\bb{s}$ quantifies the non-equilibrium spin polarization per sample area. The non-equilibrium lesser Green's function $\mathcal{G}^{<}$ is defined on the Keldysh contour \cite{Kadanoff,Rammer}. In our case, the external field depends on time but does not depend on the coordinate, hence $\mathcal{G}^<({\bb{r}},t;{\bb{r}'},t')$ is a function of three variables: $\bb{r}-\bb{r}'$, $t$ and $t'$. In the absence of light $\mathcal{G}^<$ is reduced to the equilibrium Green's function $G^<(t-t')$ which in frequency domain takes the form
\be
G^<(\omega)=-f(\omega)\lt(G^R(\omega)-G^A(\omega)\rt),
\e
where $f(\omega)=\lt[1+\exp\lt((\omega-\ep_\textrm{F})/T\rt)\rt]^{-1}$ is the Fermi distribution function with the chemical potential $\ep_\textrm{F}$ and the temperature $T$.

%%%%%%%%%%%%%%%%%%%%%%%%%%%%
%%%% fig:field
%%%%%%%%%%%%%%%%%%%%%%%%%%%%
%%%%%%%%%%%%%%%%%%%%%%%%%%%%%

The Green's function $\mathcal{G}^{<}$ is calculated perturbatively in $\hat{U}$ with the help of the Dyson equation in the Keldysh space. The first and second order contributions to $\mathcal{G}^{<}$ are given by
\beml
\label{Dyson-equation}
\beq
\label{first_order}
\delta^{(1)}\mathcal{G}^<&=& G^{R} \hat{U} G^{<}+G^{<} \hat{U} G^{A},\\
\delta^{(2)}\mathcal{G}^{<}&=& G^{R} \hat{U} G^{R} \hat{U} G^{<}+G^{A} \hat{U} G^{<} \hat{U} G^{R}\n\\
&&+G^{<} \hat{U} G^{A} \hat{U} G^{A},\label{nonequilibrium-spin}
\eq
\eml
where all products assume integral convolution in space and time arguments.

The Fourier transform of the perturbing potential reads
\be
U(\bb{p},\omega)=u_{\bb{p}}\delta(\omega-\Omega)+u\h_{\bb{p}}\delta(\omega+\Omega),
\e
where the momentum-dependent quantity $u_{\bb{p}}$ takes the form of the following matrix in the spin space
\be
u_{\bb{p}}=\frac{e \mathcal{E}_0}{i  \Omega\sqrt{1+\lambda^2}}
\bpm \frac{p}{m} e^{i \lambda \phi_p} & -i \alpha_\textrm{R} (1+\lambda)  \\ i \alpha_\textrm{R} (1-\lambda)  & \frac{p}{m} e^{i \lambda \phi_p} \epm,
\e
with the 2D momentum $\bb{p}=p(\cos{\phi_p},\sin{\phi_p})$.

In what follows we are interested in the time-independent ({\em{dc}}) contribution from conduction electrons to the magnetic field $\bb{\mathcal{H}}$. To obtain this contribution we average the induced spin polarization over time on the scale of the inverse frequency.  The resulting dc contribution is absent in first order with respect to the amplitude of light, the Eq.~(\ref{first_order}). In the second order perturbation theory of the Eq.~(\ref{nonequilibrium-spin}) the non-equilibrium spin polarization is given by the general formula
\begin{widetext}
\beml
\label{effective-field}
\beq
\label{line1}
s_a&=&- \frac{1}{2i} \int \! \frac{d^2{\bb{p}}}{(2\pi)^2} \int\! \frac{d\omega}{2\pi}\Big\la
\tr \big[\sigma_a  G^{R}_{\bb{p}}(\omega) \hat{u}_{\bb{p}} \big(G^R_{\bb{p}}(\omega-\Omega)-G^A_{\bb{p}}(\omega-\Omega)\big) \hat{u}\h_{\bb{p}} G^{A}_{\bb{p}}(\omega)\big]\big(f(\omega-\Omega)-f(\omega)\big)\\
\label{line2}
&&-\tr\big[\sigma_a  G^{R}_{\bb{p}}(\omega) \hat{u}\h_{\bb{p}} \big(G^R_{\bb{p}}(\omega+\Omega)-G^A_{\bb{p}}(\omega+\Omega)\big) \hat{u}_{\bb{p}} G^{A}_{\bb{p}}(\omega)\big]\big(f(\omega)-f(\omega+\Omega)\big)\\
\label{line3}
&&-2i\im \tr\big[\sigma_a  G^{R}_{\bb{p}}(\omega) \hat{u}_{\bb{p}} G^{R}_{\bb{p}}(\omega-\Omega) \hat{u}\h_{\bb{p}} G^{R}_{\bb{p}}(\omega)+
\sigma_a  G^{R}_{\bb{p}}(\omega) \hat{u}\h_{\bb{p}} G^{R}_{\bb{p}}(\omega+\Omega) \hat{u}_{\bb{p}} G^{R}_{\bb{p}}(\omega)\big]f(\omega)\Big\ra,\quad
\eq
\eml
\end{widetext}
that we analyze in the next Section in the diffusive approximation.

\section{Light-induced effective magnetic field}\label{sec:myfield}

\subsection{Born approximation}
The effective magnetic field given by Eq.~(\ref{effective-field}) is calculated in the limit of weak disorder $\gamma=(\gamma_++\gamma_-)/2\ll M, \Omega$. We also require $\gamma\ll \alpha_\textrm{R} p_F^-$, where $p_F^-$ is the Fermi momentum of the smaller of the two Fermi surfaces. The latter condition is equivalent to $\gamma\ll \sqrt{2\Delta(\ep_\textrm{F}-M)}$. We perform the calculation in the leading order with respect to the small parameter $\gamma/M$. For circularly polarized light, the symmetry of the problem dictates that $s_x=s_y=0$. The result for $s_z$ is, however, of the order of $M/\gamma$, that is formally diverging in the clean limit $\gamma\to 0$.  For linearly polarized light we can expect finite in-plane components of the effective magnetic field. We estimate, however, that for $x$-polarized light $s_x \propto \mathcal{O}(\gamma^0)$ and does not depend on the scattering rate while $s_y \propto \mathcal{O}(\gamma^1)$ is vanishingly small.

Below we focus on the largest component of the non-equilibrium spin polarization  $s_z$ that is of the order of $\mathcal{O}(\gamma^{-1})$. In this order we can completely disregard the contribution of the Eq.~(\ref{line3}) given by the product of retarded Green's functions, since it only contributes in the order $\mathcal{O}(\gamma^{1})$. Similarly, we can completely disregard the contributions resulting from the diamagnetic term $e^2 A(t)^2/2m$ in the Hamiltonian.

Thus, we are dealing with the calculation of $s_z$ from Eqs.~(\ref{line1},\ref{line2}) that is schematically represented by the diagrams depicted in Fig.~\ref{fig:field}(c). In order to perform the calculation we employ the self-consistent Born approximation with respect to disorder potential that amounts to the calculation of two quantities: the disorder-averaged Green's function $\bar{G}^R_{\bb{p}}(\omega)=\lt[\omega-H_0(\bb{p})+\Sigma\rt]^{-1}$ , where $\Sigma$ is a matrix self-energy, and the vertex $\Gamma_z$ depicted in the Fig.~\ref{fig:field}(b). We note that such ladder approximation remains valid as far as the resulting effective field is of the order of $M/\gamma$. To compute, for example, the $x$-component of the field that appears in the next order (for the linearly polarized light) one needs to add the diagrams with a single crossing of impurity lines \cite{Ado1,Ado2}. Due to the condition $\Omega\gg \gamma$ we do not need to consider diffusion corrections (vertex corrections) to electron-photon interaction vertices.

To compute the leading order result it is sufficient to take the self-energy $\Sigma$ in the first Born approximation illustrated diagrammatically in Fig.~\ref{fig:field}(a). For $\omega>M$ the straightforward calculation gives
\be
\im\Sigma=\gamma+\sigma_z\delta,\quad\; \gamma_\pm=\frac{1}{2}m\,\nimp V_\pm^2=\gamma\pm\delta,
\e
where $\delta$ is the imbalance between the spin up and the spin down scattering rates. Here we ignore the real part of the self-energy since it only renormalizes the exchange term. The averaged Green's function is, then, obtained as
\be
\label{GR}
\bar{G}^R_{\bb{p}}(\omega)=
\frac{\ep+i\gamma-\xi+(M-i\delta)\sigma_z+\sqrt{2\Delta\xi}\sigma_\phi}
{(\ep-x_+)(\ep-x_-)},
\e
where $\xi=p^2/2m$, $x_\pm=\xi-i\gamma\pm\sqrt{M^2+2\xi\Delta-2iM\delta}$,
and $\sigma_\phi=\sigma_y\cos\phi_p-\sigma_x\sin\phi_p$.

Thus, the equation on the vertex $\Gamma_z$, see Fig.~\ref{fig:field}(b), takes the form
\be
\label{eq}
\Gamma_z=\sigma_z+\nimp \int\frac{d^2\bb{p}}{(2\pi)^2}\lt[\hat{V} \bar{G}^A_{\bb{p}}\Gamma_z \bar{G}^R_{\bb{p}} \hat{V}\rt].
\e
This equation is solved by noting that
\beq
&&\nimp \int\frac{d^2\bb{p}}{(2\pi)^2}\lt[\hat{V} \bar{G}^A_{\bb{p}}(\omega)\sigma_z \bar{G}^R_{\bb{p}}(\omega) \hat{V}\rt]\\
&&=\frac{2 M^2 \sigma_z} { \big(\frac{\gamma_+}{\gamma_-}(\omega+M) +\frac{\gamma_-}{\gamma_+}(\omega-M)\big)\Delta+2 \lt(\omega \Delta  +M^2\rt)},\qquad\n
\eq
where $\omega>M$ and the integral over momentum $\bb{p}$ is calculated to the leading order in $\gamma/M$. The resulting solution of the Eq.~(\ref{eq}) reads
\be
\label{vertex2}
\Gamma_z=\lt[1+\frac{M^2 (1-\delta^2/\gamma^2)}{2\Delta(\omega+M\delta/\gamma)}\rt]\sigma_z,
\e
where the formal divergency at $\omega=-M\delta/\gamma$ lays outside the applicability range ($\omega>M$) of the expression. It is worth noting that the corresponding vertex corrections for the other two spin-operators $\sigma_{x}$ and $\sigma_y$ are negligible for $\Delta\ll M$\cite{Garate} in contrast to the vertex $\Gamma_z$.

In order to average $s_z$ over disorder in the diffusive approximation we have to replace the Pauli matrix $\sigma_z$ in the Eq.~(\ref{effective-field}) by the full vertex $\Gamma_z$ and the Green's function $G^{R,A}$ by the corresponding averaged Green's functions $\bar{G}^{R,A}$ from the Eq.~(\ref{GR}). As was already noted the contribution of the last term given by the Eq.~(\ref{line3}) is negligible. The rest of the calculation amounts to the averaging over the angle $\phi_p$ and two integrations over $\omega$ and $\xi$. These integrations are straightforward in the limit $\gamma\ll M$. In this limit, the product $\bar{G}^R(\omega)\bar{G}^A(\omega)$ and the difference $\bar{G}^R(\omega\pm\Omega)-\bar{G}^A(\omega\pm\Omega)$ are proportional to the corresponding delta-functions that make the calculation trivial.

\subsection{The main result}

The result of the calculation in the leading order with respect to the small parameter $\gamma/M$ reads
\beml
\label{result0}
\beq
s_z\!\! &=& \!\!\frac{(e\mathcal{E}_0)^2}{8\Omega^2} \frac{M}{\gamma}\!
\lt[\frac{\Omega^2}{4M^2}+ \frac{\lambda\Omega}{M} +1\rt]\!
R_\Omega\lt(f_- -f_+\rt),\qquad
\\
\label{ROmega}
R_\Omega \!\!&=&\!\!
M^2\frac{\tfrac{1}{4}\Omega^2-M^2-4\Delta^2+4M\Delta\delta/\gamma}{\lt(\tfrac{1}{4}\Omega^2-M^2+2M\Delta\delta/\gamma\rt)^2-\lt(\Omega\Delta\rt)^2},
\\
\label{Ffunctions}
f_\pm\!\!&=&\!\!f\lt(\frac{\tfrac{1}{4}\Omega^2-M^2}{2\Delta} \pm \tfrac{1}{2}\Omega\rt),
\eq
\eml
where the difference of the Fermi functions $(f_{-} - f_{+})$ singles out the frequency window that, at zero temperature, corresponds precisely to the kinematic condition of the Eq.~(\ref{condition}). Thus, we see that the result for $s_z$ in the order $M/\gamma$ is only due to the direct resonant optical transitions between spin-split sub-bands.

The expression for $R_\Omega$ defined by the Eq.~\eqref{ROmega} is diverging at the frequency $\Omega=2\lt(\Delta+\sqrt{M^2+\Delta^2-2 M \Delta \delta / \gamma}\rt)$. This divergence lays outside the applicability range of the formula (which is $\Omega > 2(M+2\Delta)$) and corresponds to the formal divergence in the vertex correction (\ref{vertex2}). Since $R_\Omega>0$ in the entire applicability range of the Eq.~(\ref{result0}) we conclude from Eq.~(\ref{verygeneral}) that the effective field is opposite to magnetization direction irrespective of light polarization for $g>0$ while it is in the direction of magnetization for $g<0$. Thus, both linearly as well as circularly polarized laser pulses can, in principle, lead to magnetization dynamics \cite{Gridnev}. It is, however, evident from Eq.~(\ref{result0}) that in our geometry the total non-equilibrium spin polarization $s_z$ is always larger for the case of $M$ and $\lambda$ having the same sign.

\subsection{ICM and IF contributions}

The total non-equilibrium spin polarization induced by polarized light from Eq. (\ref{result0}) can be decomposed into the sum of helicity-independent, ICM, and helicity-dependent, IF, contributions $s_z=s_\textrm{ICM}+s_\textrm{IF}$, where
\beml
\label{final}
\beq
\label{ICMc}
s_\textrm{ICM} &=& \frac{(e\mathcal{E}_0)^2}{8\Omega^2} \frac{M}{\gamma}\lt(\frac{\Omega^2}{4M^2}+1\rt)R_\Omega, \\
s_\textrm{IF} &=& \lambda\frac{(e\mathcal{E}_0)^2}{8\Omega\gamma} R_\Omega.
\label{IF}
\eq
\eml
Here we replaced $(f_--f_+)$ by $1$ assuming that $\Omega$ yields the constraint of Eq. (\ref{condition}). The helicity dependent part of the effect $s_\textrm{IF}$ is an even function of $M$, while the the helicity-independent contribution $s_\textrm{ICM}$ is an odd function of the magnetization. Thus, in the absence of magnetization, $M=0$, the ICM contribution to the effective field vanish identically. The IF contribution in this limit is still finite but negligible and can be estimated as
\be
\label{IF-no-magn}
\lt.s_\textrm{IF} \rt|_{M=0} = \lambda \gamma \frac{(e\mathcal{E}_0)^2}{2\Omega^3}.
\e
for frequencies within the window of Eq.~(\ref{condition}). We would like to stress that the resonant IF and ICM fields given by Eqs.~(\ref{final}) are orders of magnitude larger than the IF field described previously by Edelstein in the absence of magnetization \cite{Edelstein}. We find that the effective magnetic field in ferromagnetic systems due to direct optical transitions is proportional to the scattering time $\tau$ in the high frequency regime $\Omega \tau \gg 1$, while in paramagnetic systems the effective field is inversely proportional to the scattering time. As a result the value of the effective field in ferromagnetic metals is large and can reach several Tesla.

Quite generally the IF component of the field, which is also referred to as the inverse Faraday effect, is parallel to the propagation direction of the circularly polarized light. The ICM field component is, however, parallel to the magnetization direction of the sample. In our setup these two directions are equivalent and the in-plane components of the effective field are vanishing by symmetry, $s_x=s_y=0$, for the case of circularly polarized light.

From Eqs.~(\ref{final}) we immediately obtain the ratio
\be
\label{prediction}
\frac{s_\textrm{IF}}{s_\textrm{ICM}}  = \lambda \frac{\Omega M}{\tfrac{1}{4}\Omega^2+M^2},
\e
which depends solely on the sample magnetization and the frequency of light. The prediction of Eq.~(\ref{prediction}) can be, therefore, tested in the experiment. We stress that the ratio $s_\textrm{IF}/s_\textrm{ICM}$ decays monotonously as the function of $\Omega$ in the frequency range set by Eq.~(\ref{condition}), since the latter always assumes $\Omega > 2M$. Thus, the maximal value of the total non-equilibrium spin polarization $s_z$ is achieved for the smallest frequency allowed by the constraint (\ref{condition}) and for $\lambda=\mathrm{sgn}[M]$ (see also Fig.~\ref{fig:results}).

It is worth noting that the calculation of the vertex correction in Eq.~(\ref{vertex2}) corresponds to the electron diffusion that is a relatively slow process. Thus, the results of Eqs.~(\ref{result0}) and (\ref{final}) are qualitatively accurate only for the light pluses that last longer than the electron transport time, which is typically of the order of picosecond in clean systems. However, the effect of light described by Eq.~(\ref{effective-field}) is effective already on very short time scales. For an estimate of the effect for femto-second light pulses in dirty samples one should simply consider $\Gamma_z=\sigma_z$ which corresponds to $R_\Omega\to R'_\Omega$ with
\be
R^{'}_\Omega=\frac{2M^2\lt(\tfrac{1}{4}\Omega^2-M^2-4\Delta^2\rt)}{\Omega\lt(\tfrac{1}{2}\Omega-2\Delta\rt)\lt(\tfrac{1}{4}\Omega^2+\Omega\Delta-M^2\rt)},
\e
where we have taken $\delta=0$ for compactness. It is easy to see that the expression for $R'_\Omega$ is of the same order as  $R_\Omega$ hence we may expect equally strong effective magnetic field on the short time scales.

\section{Optical conductivity}\label{sec:conductivity}

\subsection{The Kubo formula}
The resonant optical transition between spin-split sub-bands leads naturally to an energy dissipation in the system. This may limit the applicability of our results on long time-scales due to heating effects. However, if the duration of the light pulse is shorter or comparable to the electron scattering time, the effect described in the previous section must play an important role while the energy absorbed by the system is still small. Sub-picosecond light pulses may definitely fulfill these constraints.

In order to quantify the light absorption we analyze the optical conductivity tensor $\sigma_{ab}(\Omega)$ at the frequency $\Omega$ \cite{Rashba-conductivity4}. The real part of the conductivity, $\re \sigma_{xx}$, represents in-phase current which induces a resistive joule heating, while the imaginary part, $\im \sigma_{xx}$, represents $\pi/2$ out-of-phase inductive current. If the inductive current is dominant the energy absorption and the related joule heating are negligible.

To simplify the further analysis we let $\delta=0$ i.\,e. consider the spin-independent disorder scattering rate. We also assume the high frequency limit $\gamma \ll \Omega$. The latter condition allows us to ignore vertex corrections and use disorder-averaged Green's functions of the Eq.~(\ref{GR}) in the Kubo formula for conductivity. In the imaginary time representation the conductivity tensor taken at imaginary frequencies is given by
\be
\label{optical-conductivity}
\sigma_{ab}(i\nu_m)=\frac{T}{\nu_m}\s_n\!\int\!\!\frac{d^2{\bb{p}}}{(2\pi)^2}
\tr\lt[j_a\bar{G}_{\bb{p},n} j_b \bar{G}_{\bb{p},n+m}\rt],
\e
where $T$ is the temperature, $\nu_m=2\pi Tm$ is the bosonic Matsubara frequency, and $m, n$ are the integers. It is convenient to represent the disorder-averaged Matsubara Green's function as a sum of two terms that are related to different sub-bands,
\be
\label{Matsubara}
\bar{G}_{\bb{p},n}=\frac{1}{2}\s_{s=\pm}\lt(1-s\frac{\bb{\sigma} \cdot \bb{B}_{\bb{p}}}{B_{\bb{p}}}\rt)\frac{1}{i(\omega_n+\gamma)-\ep^s_{\bb{p}}},
\e
where $\omega_n=\pi T(2n+1)$ is the fermionic Matsubara frequency and $\bb{B}_{\bb{p}}=(-\alpha_\textrm{R} p_y,\alpha_{\textrm{R}} p_x,M)$. It is easy to see from symmetry properties of our model that $\sigma_{xx}=\sigma_{yy}$ and $\sigma_{xy}=-\sigma_{yx}$. The conductivity at a real frequency $\Omega$ is obtained from the Eq.~(\ref{optical-conductivity}) by the analytic continuation.

\subsection{Longitudinal conductivity}

We focus first on the longitudinal optical conductivity $\sigma_{xx}(\Omega)$. With the help of the Eq.~(\ref{Matsubara}) we decompose the latter into the sum of intraband and interband contributions, $\sigma_{xx}=\sigma_{xx}^{\textrm{intra}}+\sigma_{xx}^{\textrm{inter}}$. The intraband component of the conductivity does not involve any direct optical transitions and can be evaluated at zero temperature as
\beq
\sigma^{\mathrm{intra}}_{xx}(\Omega)
&=&\frac{e^2}{\gamma-i\Omega}\int\frac{d^2{\bb{p}}}{(2\pi)^2}
\sum_{s=\pm}\frac{p_x^2 (\Delta-s B_{\bb{p}})^2}{m^2 B^2_{\bb{p}}} \frac{\pa f(\ep^s_{\bb{p}})}{\pa \ep_\textrm{F}}\n\\
&=&\frac{2 e^2}{\pi}\,\frac{\ep_\textrm{F}+\Delta}{\gamma-i \Omega},
\label{intra-optical-conductivity}
\eq
which can be recognised as the Drude conductivity. For $\Omega\gg\gamma$ the imaginary part of the intraband conductivity is proportional to $(\ep_\textrm{F}+\Delta)/\Omega$, while the real part is suppressed by the large factor $\Omega/\gamma$. Thus the contribution of the intraband term to the energy absorption is negligible.

The interband contribution to the longitudinal conductivity is given by,
\beq
\sigma^{\mathrm{inter}}_{xx}(\Omega)
&=&\frac{e^2\alpha^2_\textrm{R}}{i\Omega}\!\!\int\!\!\frac{d^2{\bb{p}}}{(2\pi)^2} \frac{\alpha_\textrm{R}^2 p_y^2+M^2}{B^2_{\bb{p}}}\lt[f(\ep^+_{\bb{p}})-f(\ep^-_{\bb{p}})\rt]\n\\
&\times& \lt(\frac{1}{\Omega-2B_{\bb{p}}+i0}-\frac{1}{\Omega+2B_{\bb{p}}+i0}\rt),
\label{inter-optical-conductivity}
\eq
where we have taken the limit $\gamma\to 0$.

At zero temperature the integrals in the Eq.~(\ref{inter-optical-conductivity}) are easily taken with the result,
\beml
\beq
\label{inter-conduc-23a}
\re\sigma^{\mathrm{inter}}_{xx}&=& \frac{e^2}{2}\lt(1+\frac{4 M^2}{\Omega^2}\rt) \Theta(1-\eta^2),\\
\label{inter-conduc-23b}
\im\sigma^{\mathrm{inter}}_{xx}&=&\frac{e^2}{2\pi}\lt[\frac{1}{8}\lt(1+\frac{4 M^2}{\Omega^2}\rt)\ln\lt|\frac{\eta-1}{\eta+1}\rt|-\frac{\Delta}{\Omega}\rt],\qquad\;
\eq
\eml
where $\Theta(x)$ stands for the Heaviside step-function and we introduce the parameter
\be
\eta=\frac{1}{\Omega\Delta}\lt(\tfrac{1}{4}\Omega^2-2\ep_\textrm{F}\Delta-M^2\rt).
\e
We see that in the limit $\gamma\to 0$, the real part of the inter-band conductivity (\ref{inter-conduc-23a}) is finite only for $|\eta|<1$. This condition is identical to the kinematic condition of the Eq.~(\ref{condition}) for the direct optical transitions. In this frequency range the imaginary part of the conductivity is generally smaller than the real one.

Thus, the absorption of light is indeed associated with the direct optical transitions. In the absence of magnetization $M=0$, we naturally reproduce the well-known results for optical conductivity of 2DEG-Rashba systems \cite{Rashba-conductivity1,Rashba-conductivity2,Rashba-conductivity3,Rashba-conductivity4}.

\subsection{Anomalous Hall conductivity}
For the completeness we also provide the expression for the transverse component of the optical conductivity, the so-called {\em ac} anomalous Hall conductivity, in the limit $\gamma\to 0$, given by
\be
\label{ac-anomalous-Hall}
\sigma_{xy}(\Omega)= \frac{e^2 M}{2\pi\Omega}\ln\lt|\frac{\eta-1}{\eta+1}\rt|+i\frac{e^2 M}{2\Omega}\Theta(1-\eta^2).
\e
The real part of the {\em ac} anomalous Hall conductivity defines the so-called Faraday rotation angle, the rotation of the plane of light polarization traveled through a magnetic medium, which is zero in the absence of magnetization \cite{Faradayrotation3,conductivity-absorption,Faradayrotation4,Faradayrotation1,Faradayrotation2}. At zero frequency $\Omega=0$ we naturally recover the intrinsic value of the {\em dc} anomalous Hall conductivity $\sigma^{\textrm{int}}_{xy}[e^2/2\pi]=2M\Delta/(M^2+2\varepsilon_F\Delta)$ \cite{sinova,Ado2}.

\section{Conclusion and discussion}
\label{sec:conclude}

\begin{figure}[bt]
\includegraphics[width=\columnwidth]{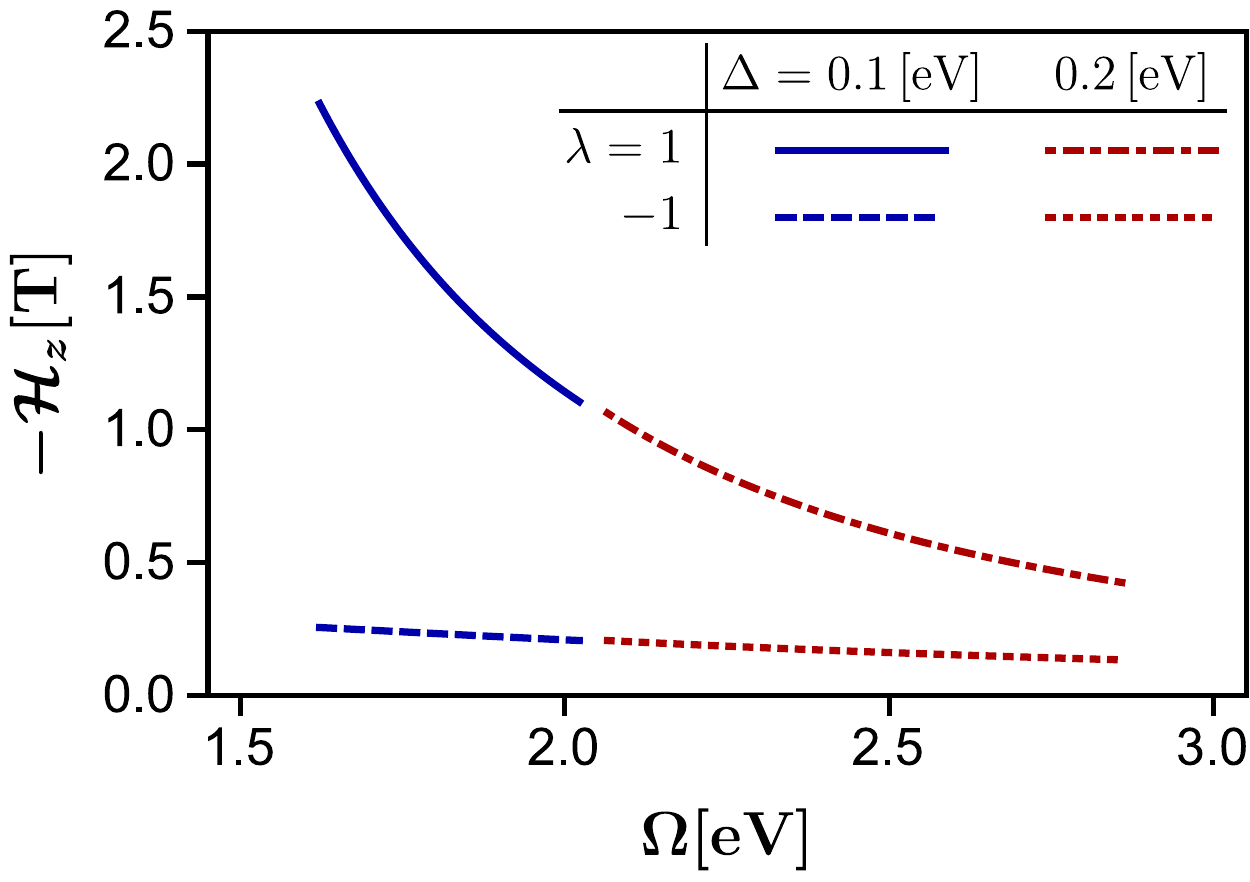}
\caption{The results for $z$-component of the effective magnetic field in the units of Tesla calculated from Eqs.~(\ref{verygeneral},\ref{result0}) for circularly polarised light with the helicity $\lambda$ and for two values of the parameter $\Delta=m\alpha_\textrm{R}^2$. We choose here: $\mathcal{E}_0=10^9$\,Vm, $\ep_F=3.3$\,eV, $M=0.3$\,eV, $\gamma_+=2\gamma_-$, $\gamma_-=1$\,meV, $d=0.4$\,nm. The effect is restricted to a finite frequency window of Eq.~(\ref{condition}) that essentially depends on $\Delta$.}
\label{fig:results}
\end{figure}

We analyzed the effective magnetic field induced by polarized light in a 2D ferromagnet with Rashba spin-orbit coupling. We assumed that the light propagates along the magnetization direction that is perpendicular to the plane of the ferromagnet. In an interval of frequencies set by the kinematic condition (\ref{condition}) we predict a greatly enhanced effect due to direct optical transitions between spin-split sub bands. We find that in ferromagnetic systems the sign of the total light-induced effective magnetic field is not proportional the light helicity. The field can, however, be decomposed to the helicity dependent term (IF effect) and a helicity independent term (ICM effect). This is different with non-magnetic systems where the overall sign of the effective field is set by the light helicity, the IF effect. The actual direction of the light-induced magnetic field in ferromagnet depends also on the sign of the effective g-factor. If the effective g-factor of charge carriers were positive the total light-induced effective magnetic field direction would be always opposite to the initial magnetization direction of the sample irrespective of the light polarization (in case of the linearly polarized light, a small transverse component of the field also emerges parallel to the polarization vector). Thus, whether the effective magnetic field leads to a magnetization reversal is material dependent phenomenon. The magnitude of the effect, given by the Eqs.~(\ref{result0}, \ref{final}) and illustrated in the Fig.~\ref{fig:results}, is proportional to a large parameter $M/\gamma$ where $M$ is the exchange energy and $\gamma$ is the scattering rate. The effect takes place also on short time-scales, i.\,e.\, for femtosecond time pulses. Both effects, IF and ICM, can be classified as the optical rectification \cite{Oppeneer,opticalrectification} in magnetic systems where an ac electric field induces a dc magnetic field.

In contrast, the effective magnetic field in the absence of magnetization is strongly suppressed \cite{Edelstein,Tatara3}, see Eq.~(\ref{IF-no-magn}). The effect in this case is proportional to a small parameter $\gamma/\Omega$, where $\Omega$ is the frequency of light, and can only be induced by a circularly polarized light.

The strength of spin-orbit interaction defines the frequency window of Eq.~(\ref{condition}) in which the effect can be observed. Since the effect is facilitated by resonant optical transitions it is always accompanied by some joule heating. The energy absorbed depends crucially on the time of the light pulse and the sample quality.

%%%%%%%%%%%%%%%%%%%%%%%%%%%%
%%%% fig:results
%%%%%%%%%%%%%%%%%%%%%%%%%%%%
%%%%%%%%%%%%%%%%%%%%%%%%%%%%%

In Fig.~\ref{fig:results} we illustrate our main results, which are given by Eqs.~(\ref{verygeneral}) ans (\ref{result0}), using typical parameters: $\mathcal{E}_0=10^9$\,Vm, $\ep_F=3.3$\,eV, $M=0.3$\,eV, $\gamma_+=2\gamma_-$, $\gamma_-=1$\,meV, $d=0.4$\,nm. The effective field is shown for two different values of spin-orbit interaction strength: $\Delta=0.1$\,eV and $0.2$\,eV such that the condition of Eq.~(\ref{condition}) selects the frequency windows $\Omega\in (1.62, 2.02)$\,eV and $\Omega\in (2.07, 2.87)$\,eV, correspondingly. We see that the effective magnetic field can exceed $1$ T for the case of circular light polarization with the helicity $\lambda=\mathrm{sgn}[M]$. We also see that the IF contribution is indeed larger for lower values of $\Omega$ in the frequency window of Eq.~(\ref{condition}) as expected from Eq.~(\ref{prediction}). Our results show that we get an effective magnetic field of between $0.2$ T to $2.2$ T within realistic parameter rage for Co/Pt bilayers . Very recently it was experimentally reported that the light induced effective magnetic field is about $0.2$ T in such heterostructures \cite{Huisman}.

The effect of spin-orbit interaction on inverse Faraday effect has been studied recently by several authors. In the Ref.~\onlinecite{Tatara2} it was shown that the IF effect in disordered metals and in the terahertz regime is linearly proportional to both the frequency of light and the extrinsic spin-orbit interaction strength. In clean metals, in the presence of the Rashba spin-orbit interaction, the effect was found to be of the second order in $\alpha_\textrm{R}$ in the terahertz regime \cite{Tatara3}. On the other hand, the IF effect in dilute magnetic semiconductors has been argued to have an intrinsic origin due to the renormalisation of the spin-split sub-bands via the spin-dependent optical Stark effect \cite{Alireza1}. The resulting effective magnetic field in such a scenario (for both IF and ICM effects) is inversely proportional to $\Omega^2$ and is vanishing in the absence of the spin-orbit interaction. The mechanism, however, requires the light pulse duration to be larger than the electron spin-flip scattering time. The latter is typically of the order of a few picoseconds \cite{Alireza1}. In contrast, the light-induced magnetic field due to direct optical transitions between the spin-split sub-bands, which we considered above, is effective already on ultra-short time scales and is of particularly large amplitude.

Our results are also relevant in the context of all-optical magnetization reversal \cite{Kimel2005,Stanciu2007}. It was shown in Refs.~\onlinecite{linear-reversal,chantrell} that for temperatures in a vicinity of the Curie temperature, the linear magnetization switching can occur provided the magnetic field applied is larger than a critical value that is typically of the order of few Tesla. The authors also showed that at lower temperatures the magnetization may switch via the precessional reversal in similar magnetic fields but on longer time scales. Our calculation shows that the light pulse in the kinematic frequency range of the Eq.~(\ref{condition}) induces both the joule heating as well as the large {\em dc} effective field on the scale of $1$ Tesla. Thus, the mechanism described may be optimized for all-optical magnetization reversal \cite{IFE-review0}.

In conclusion we derived the effective magnetic field due to direct optical transitions between the spin-split sub-bands in the presence of both magnetization and Rashba spin-orbit coupling. The effect can be used to initiate the ultrafast magnetization dynamics for all-optical magnetization reversal.

\section*{Acknowledgements}
We would like to thank I.\,A.\,Ado, G.\,E.\,W.~Bauer, R.\,A.~Duine, M.\,I.~Katsnelson, A\,Kimel, J.\,H.~Mentink and Th.~Rasing for discussions. This work was supported by Dutch Science Foundation NWO/FOM 13PR3118 and by EU Network FP7-PEOPLE-2013-IRSES Grant No 612624 "InterNoM".

\end{document}